\documentclass[preprint,aps,showpacs,preprintnumbers,amsmath,amssymb]{revtex4}
\usepackage{graphicx}
\usepackage{dcolumn}
\usepackage{bm}
\usepackage{color}

\begin{document}
\title{Coulomb oscillations of the Fano-Kondo effect and zero bias anomalies in the double dot meso-transistor}
\author{A. Aldea$^{1,2}$, M. \c Tolea$^{1}$, I.V. Dinu$^{1}$}
\affiliation{\hskip-1.4cm $^1$ National Institute of Materials
Physics, POB MG-7, 77125
 Bucharest-Magurele, Romania. \\
$^2$ Institute of Theoretical Physics, Cologne University, 50937
Cologne, Germany.}
 \begin{abstract}
We investigate theoretically the transport properties of the side-coupled
double quantum dots in connection with the experimental
        study  of Sasaki {\it et al.} Phys.Rev.Lett.{\bf 103}, 266806 (2009).
The novelty of the set-up consists in connecting
          the Kondo dot directly to the leads, while the side dot provides
	  an interference path which affects the Kondo correlations.
        We analyze the oscillations of the source-drain current
	due to the periodical Coulomb blockade of the many-level
	side-dot at the variation of the gate potential applied on it.
        The Fano profile of these oscillations may be controlled by the
        temperature, gate potential and interdot coupling.
The non-equilibrium conductance of the
 double dot system exhibits zero bias anomaly which, besides
        the usual enhancement, may show also a suppression
         (a dip-like aspect) which  occurs
        around the Fano {\it zero}. In the same region,
	the weak temperature dependence of the conductance indicates
	the suppression of the Kondo effect.
	Scaling properties of the non-equilibrium
        conductance in the Fano-Kondo regime are discussed. Since the
	SIAM Kondo temperature is no longer the proper scaling parameter,
	we look for an alternative specific to the double-dot.
        The extended Anderson model, Keldysh formalism and equation
        of motion technique are used.

\end{abstract}
\maketitle
\section{ Introduction }
The problem we address enter the family of effects resulting
from the interplay of interference processes, Kondo
correlations and Coulomb blockade in complex quantum dot
systems. The specific mesoscopic set-up we study (shown in
Fig.1) is a side-coupled double quantum dot  with the small dot
connected to external leads and also to a larger lateral dot.
The set-up resembles the transistor structure, where the
lateral big dot plays the role of the 'basis' and controls the
current flowing from the left lead (the source) to the right
lead (the drain).

The small dot is a "Kondo dot" containing one degenerate level
$E_d$ which is at most
single occupied due to the strong Hubbard repulsion, and can be tuned
 in different regimes: Kondo, mixed valence or
empty level. The side-dot contains many levels, and the
access of the charge carriers into this dot is controlled by the
Coulomb blockade  which can be periodically switched on/off by a
continuous variation of the gate potential $V_g$. In this way  the
interference conditions  in the device  can be changed  periodically
resulting in a periodical modulation of the source-drain
conductance. The  modulation will have a Fano shape which depends on
the inter-dot coupling and the regime of the small (central) dot,
which may work in Kondo or non-Kondo mode.

We have two motivations for considering a
"large" side-coupled dot. First, because the electronic correlations
are much weaker in a large dot, and, this way, the side-coupled dot
basically provides just an interference path as considered by Sasaki
{\it et al.} \cite{Sasaki} in a recent experiment that motivated our work.
The interesting physics arises from the fact that the interference
occurs during the Kondo tunneling and the Kondo correlations are
also affected. The second
reason is that a large dot has small level spacing which can
influence the shape of the Fano resonances.

\begin{figure}[htb]
\includegraphics[angle=-00,width=0.40\textwidth]{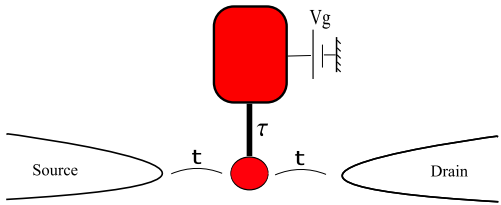}
\caption{(color online) The sketch of the transistor-like double dot system }
\end{figure}

The side-coupled double dot system has been much studied  in the
recent literature, and various aspects has been put forward.
The two-stage Kondo effect was proven if the two quantum dots are small and
the spin screening takes place in both of them \cite{Cornaglia,Chung}.
The interplay between Fano and Kondo processes, known as Fano-Kondo effect,
is specific to such a  geometry.
Wu {\it et al.} \cite{Wu} studied the simple T-shape model which  includes
a single energy level in each dot. The slave-boson mean field theory
(used at low bias  and zero temperature) indicates
the absence of the unitary limit of the linear conductance
 and also the suppression of the zero bias
(peak-type) anomaly when the side dot is coupled.
Recently, as already mentioned, Sasaki {\it et al.} \cite{Sasaki,Tamura}
investigated experimentally
the linear conductance as function of the gate on the lateral dot,
both in the Kondo and non-Kondo regime of the  dot connected to leads.  One
proves  the suppression of the Kondo effect by Fano
destructive interferences.
The finding is supported by a calculation in the slave-boson
approximation again for the  simple T-shape model,
where the interference takes place between the
resonant channel passing through the side-dot and the continuous Kondo channel (see also \cite{Liu}).
The problem is reassumed by Zitko \cite{Zitko} who suggests that the
experiments in \cite{Sasaki}  may be also understood  in the frame
of a two stage
Kondo effect if the experimental temperature  lies between the two Kondo
temperatures.
The influence of ferromagnetic leads in the presence of an exchange between
the two dots is studied in \cite{Barnas}.

The non-equilibrium regime of  quantum dot devices raises the question
of the so-called zero bias anomaly (ZBA), consisting
in a singular behavior of the differential conductance $dI/dV$ at $V=0$.
ZBAs occur in  various physical systems like semiconductor impurity bands
\cite{Shklovskii}, tunnel junctions \cite{Altshuler} or quantum dots.
For instance, the dip of the differential conductance
found in metal-insulator-metal tunneling is attributed to the dip in the
density of states of the electrodes. An elaborate theory due to
Altshuler and Aronov proves that the disorder  gives rise to such an effect
if accompanied by  electron-electron interaction. So, such anomalies
are considered as a fingerprint of the Coulomb interactions.

ZBAs are present also in  mesoscopic systems, as for instance, the
strong maximum of the differential conductance  observed in quantum dots
below the Kondo temperature;
the maximum is understood as corresponding
to the Kondo-peak in the  density of states of the dot (see for instance
\cite{Koenig,Kastner}).
The interaction is again a necessary ingredient, however, as a difference
from the case of the tunneling junction mentioned above,
the interaction occurs now inside the dot,
and the information of interest is carried by the density of states of the
dot, while the lead spectrum is described by the simple flat band model.
Since the experiment observes the suppression of the Kondo resonance due to
destructive interference \cite{Sasaki} one may ask whether the peak-type
 anomaly  might not be replaced by a dip-type anomaly under specific conditions
in  double-dot systems. This question is one of the topics of the paper.

The scaling properties of the non-equilibrium conductance is a topic of
present interest in the field, most papers addressing the single impurity
Anderson model, e.g. \cite{Ralph,Schiller,Oguri,Grobis,Scott,Majumdar}.
In this paper we perform a scaling analysis of the
nonequilibrium conductance
$G=dI/dV$  of the transistor-like device based on  double lateral dots
connected to  source and drain. We use the data collapse technique, i.e.
we look for that scaling of $\Delta G(V,T)=G(0,T)-G(V,T)$  and of
the variables $V,T,T_K$  which produces the
collapse of all numerical curves in a single one.
The specific situation consists in the dip-peak crossover which means the
change of sign  of the curvature  at the variation of the parameters
$V_g$ or $T$.
This  makes impossible the collapse of all curves in a single one, i.e.
prevents the identification of a single -universal- scaling law.
We succeed to prove  scaling properties in the peak-like regime,
the scaling parameter being the half-width of the Kondo peak for the
double dot system.

The model we adopt is an extended Anderson Hamiltonian (see eq.(1)), and the
conductance is calculated in the Keldysh formalism. Being
interested not only in the Kondo  but also in the mixed-valence regime,
we shall use the equation of motion technique, which works well also in
this case. In Sec.II we  describe the model and the formalism. Sec.III
analyzes the spectral properties of the specific double dot model. In Sec.IV
we calculate the source-drain current versus the gate potential applied on
the big dot. The oscillatory behavior and Fano aspect of the resonances are
pointed out. Sec.V proves the peak-dip crossover of the zero bias anomaly
in the transistor-like system and discusses some scaling properties of
the nonequilibrium conductance.

\section{ The model Hamiltonian and  transport formalism}
The set-up discussed in introduction will be described by the following
 Hamiltonian , which may be considered as an extended Anderson model:
\begin{equation}
 H= \sum_{k,\sigma,\alpha} (\epsilon_k-\mu_\alpha)c_{k\sigma,\alpha}^{\dagger}
 c_{k\sigma,\alpha}
 +\sum_{\sigma} E_{d} d_{\sigma}^{\dagger}d_{\sigma} +
  U_H n_{d \uparrow}n_{d \downarrow}
  + \sum_{i\sigma} E_i c_{i\sigma}^{\dagger} c_{i\sigma} +
  \frac{e^2}{2C} N^2 \nonumber
\end{equation}
\begin{equation}
   + \sum_{k,\sigma,\alpha} t_{kd}\big(c_{k\sigma,\alpha}^{\dagger}d_{\sigma}
 +h.c. \big) + \sum_{i}  \tau_{id} \big(c_{i\sigma}^{\dagger} d_{\sigma}+h.c.
\big) ,
\end{equation}
where one may identify the Hamiltonian of the leads ($\alpha=L(left),R(right)$),
the Hamiltonian of the small dot including the
Hubbard term, and that one of the big (lateral) dot which contains also the
Coulomb repulsion in the orthodox (capacitance) model;
the operator of the total occupation number
reads $N=\sum_{i\sigma} n_{i\sigma}=\sum_{i\sigma}
c_{i\sigma}^{\dagger} c_{i\sigma}$.
The last two terms  stand for the coupling of the small  dot to the leads and
to the big dot. Without loss of generality, the energy levels of the big
dot will be treated as equidistant $E_i=E_0+i \delta E$~ $ (i=0,1,2..)$ ;
We notice that the Hamiltonian (1)  reduces either to the
usual SIAM when $\tau=0$ or to the  simple T-shape model when only one level
is considered in the  lateral dot.
The term describing the Coulomb interaction in the
big dot will not be taken explicitly into the formalism, however we shall
keep in mind that, due to the Coulomb blockade, the charging energy required for
absorbing any new electron in  the big dot equals the quantity
$\Delta= U_c +\delta E;~ U_c=e^2/C$.

The charge current flowing through the device  is given in the Keldysh
formalism by the well-known formula \cite{Meir}:
\begin{equation}
I(V)=\frac{e}{\hbar}\sum_{\sigma}\int d\omega[f_L(\omega-\frac{eV}{2})-
f_R(\omega+\frac{eV}{2})]
~\Gamma_{\sigma}(\omega) [-\frac{1}{\pi}~ Im G^{r}_{dd,\sigma}(\omega)] ,
~~
\end{equation}
where the main ingredient is the retarded Green function
 $G^{r}_{dd,\sigma}$ at the site 'd' denoting the small dot. ( $V$ is the
 applied bias, $\Gamma$
 describes the coupling of the small dot to the leads, and $f_{L/R}$ are the
Fermi functions on the left/right side). If we are
not too far from equilibrium (i.e., $eV/kT<<1$), one may neglect the derivative
of $\Gamma$ and $G^{r}_{dd}$ with respect to $V$, and the differential
conductance reads (see also \cite{Schiller}):
\begin{equation}
G(V)=dI/dV= \frac{e^2}{2h }\sum_{\sigma}\Gamma_{\sigma}~
[ Im G^{r}_{dd,\sigma}(-\frac{eV}{2})+ Im G^{r}_{dd,\sigma}(\frac{eV}{2})]~
\end{equation}
The above equation says that, for the system we take into consideration, the
whole information  about the possible zero-bias anomaly of the
differential conductance is
contained in the density of states of the small dot
$DoS_d(\omega)=-\frac{1}{\pi} Im G_{dd}(\omega)$, which should be calculated in the presence of all couplings and interactions. Since the first derivative
of $ G(V)$ vanishes at $V=0$ , the zero-bias
behavior is given by the second derivative (convexity)  of the function
$G(V)$ which is proportional to
the convexity of the spectral function at the site 'd'.
Indeed,  in the absence of the spin dependence, one has:
\begin{equation}
G"(0)= -\frac{e^4}{h}\Gamma Im G"_{dd}(0).
\end{equation}

This expression shows that an eventual change of the
spectral function convexity  yields a peak-dip crossover of the
differential conductance.

For the calculation of the Green function we use the recipe proposed by
Entin-Wohlmann et al \cite{Meir05}  which is an extension to complex mesoscopic
structures of the Lacroix equation of motion approach, originally used
for SIAM \cite{Lacroix}. The further extension of the formalism to
non-equilibrium is described in \cite{TDA}.
The input quantity required by this recipe is the non-interacting self
energy $\Sigma_0(\omega)$ (calculated in the absence of the
Hubbard term in the small dot, $U_H=0$).
In order to identify this quantity in the case of the Hamiltonian (1),
we  sketch below the equations of motion of the Green functions
specific to this problem:
\begin{equation}
(\omega^{+}-E_d) G_{dd\sigma}=1+\sum_{k,\alpha}t_{dk} G_{kd\sigma}^{\alpha}
+ \sum_{i}\tau_{di}
G_{id} + U_H \ll d_{\sigma} n_{d,-\sigma}; d_{\sigma}^{\dagger}\gg
\end{equation}
\begin{equation}
(\omega^{+}-\epsilon_k+\mu_{\alpha}) G_{kd}^{\alpha}=  t_{kd} G_{dd}
\nonumber
\end{equation}
\begin{equation}
(\omega^{+}-\epsilon_i ) G_{id}=  \tau_{id} G_{dd} . \nonumber
\end{equation}
Combining the above equations, one obtains:
\begin{equation}
\big(\omega^{+}-E_d-\Sigma_0(\omega))G_{dd}(\omega)=
1+U_H \ll d_{\sigma} n_{d,-\sigma};
d_{\sigma}^{\dagger}\gg.
\end{equation}
This is a well known formula , however for our specific lateral
 double-dot model, $\Sigma_0$ includes two terms coming from the leads and
the big dot, respectively:

\begin{equation}
\Sigma_0 (\omega)=\sum_{k,\alpha} \frac{|t_{kd}|^2}{\omega^{+}-\epsilon_k+
\mu_{\alpha}}+\sum_{i}\frac{|\tau_{id}|^2}{\omega^{+}-E_i}.
\end{equation}
One notices that the two contributions are additive
and both of them acts on the  Kondo peak of the density of states of
the small dot. The second term is that one which gives rise to the
{\it  Kondo in a  box} effect, which was studied in \cite{Simon} in the case of
vanishing dot-leads coupling. In the simple description of the flat
continuous spectrum for the leads
the first term in eq.(7) equals $i\Gamma$ (the quantity
present in  the current formula eq.(2)). On the other hand,
the second term in eq.(7) is a finite  sum over the discrete spectrum of
the big dot. This contribution can be tuned by the lowest energy level
$E_0$, which  will be identified from now on as the gate potential applied
on the big dot,  $E_0 \equiv V_g$ (see Fig.1).

\section{ The spectral function $ImG_{dd}(E)$ }
The spectral properties help with  the  identification of various
possible regimes in the transport phenomena. Since the gate potential
$V_g$ applied on the big dot is the functional parameter of
double dot device that works like a meso-transistor,
 the spectral function will be studied for different values of $V_g$.
An  analytical expression is available in the
non-interacting case $U_H=0$ , showing properties which are also  useful
 for the discussion of the numerical curves obtained in the presence of the
 interaction. From eqs.(6-7) one gets:
\begin{equation}
G^0_{dd}(\omega)=\big(\omega^{+}-E_d+i\Gamma -
\tau^2\sum_i\frac{1}{\omega^{+}-E_i}\big)^{-1}
\end{equation}
and also the following expression for the imaginary part:
\begin{equation}
-Im G^0_{dd}= \frac{A^2(\omega)\Gamma}{[(\omega-E_d) A(\omega)-
\tau^2 B(\omega)]^2 + A^2(\omega) \Gamma^2},
\end{equation}
where $A(\omega)=\Pi_i(\omega-E_i)$ and $B(\omega)=dA(\omega)/d\omega$~
with $E_i=V_g + i \Delta $ (i = integer).
\begin{figure}[htb]
\includegraphics[angle=-90,width=0.45\textwidth]{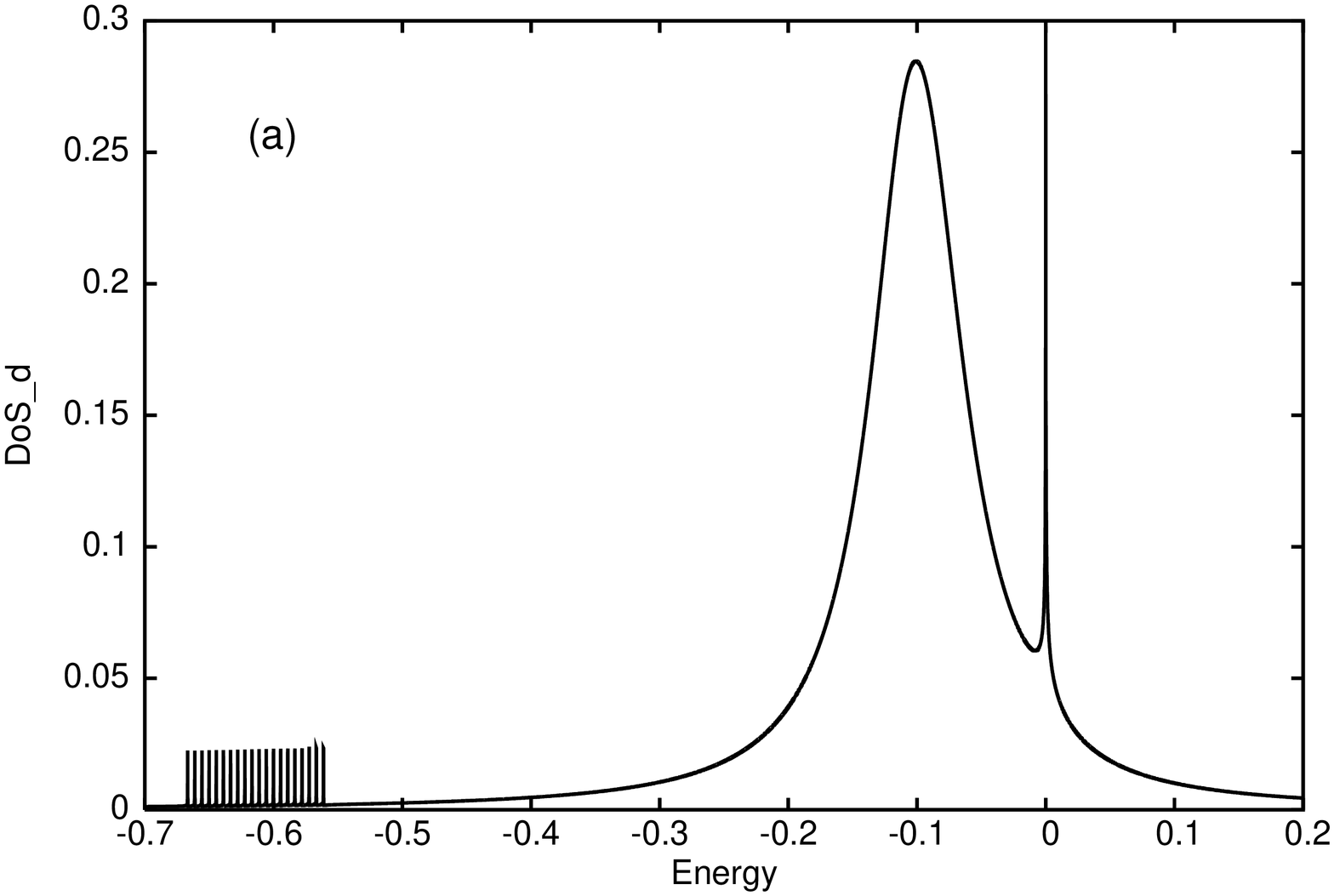}
\includegraphics[angle=-90,width=0.45\textwidth]{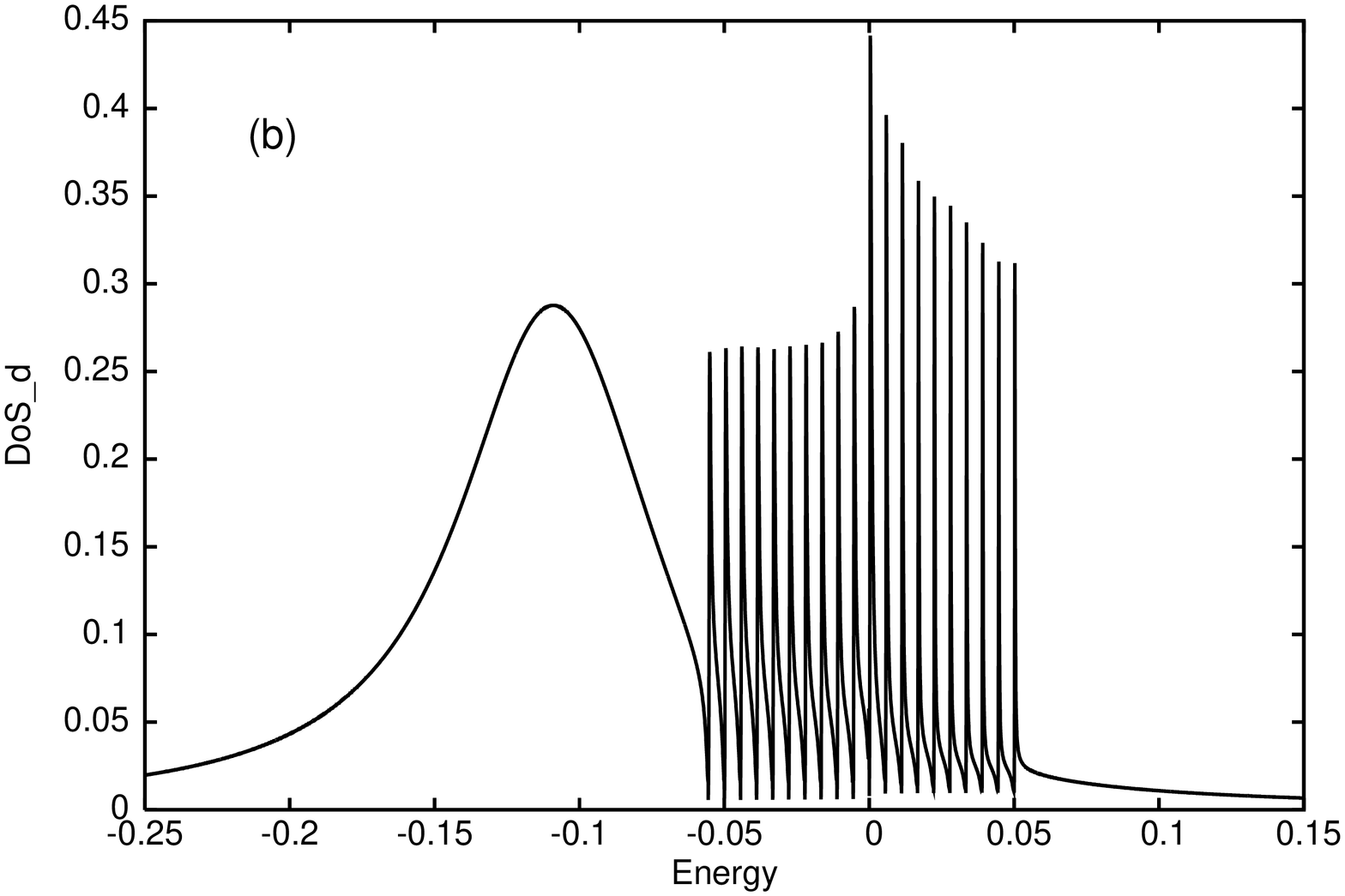}
\caption{The local density of states at the site $d$ as function of energy
for two values of the gate potential: (a) $V_g=-0.66 $
; (b) $V_g=-0.05$ ($E_d=-0.1$~; the energy unit is the half-width of the
leads band)}
\end{figure}
The eq.(10) indicates that when  $\tau\neq 0$  the
spectral function has a series of equidistant zeros at $\omega=E_i$.
This structure was noticed also by Liu \cite{Liu}.
The spectral function shows a specific structure, namely, besides the usual
d-resonance, it exhibits a series of peaks located between the mentioned
zeros .

In the interacting  case ($U_H\neq0$) , the Kondo effect arising
in the small dot superimposes its own  characteristics on the spectral
function of the double-dot system. More than this, by changing the
gate potential $V_g$ one may choose different regimes.
For instance,  when the gate is such that the bunch of levels $\{E_i\}$
is positioned much below the d-level $E_d$ the local density of states
at the d-level   $DoS_d= -\frac{1}{\pi}Im G_{dd} $,
looks as expected (see Fig 2a): the strong d-resonance carrying the  Kondo peak
at $E=E_f=0$ is the most evident,
while the oscillations coming from the  hybridization with the big dot
can  be noticed at much lower energies.
The density of states changes drastically
if we push up the gate $V_g$ such that
 the spectrum of big dot encompasses the Fermi energy;
in this case the big dot (which is now only partially filled) becomes strongly
hybridized with the small dot.
\begin{figure}[htb]
\includegraphics[angle=-90,width=0.45\textwidth]{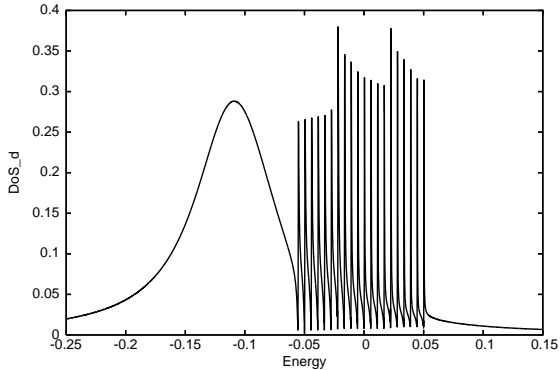}
\caption{The same as in Fig.2b in the presence of a bias $V=0.04$ }
\end{figure}

In the absence of the leads, the changes induced by the presence of the
big dot to the density of states at the site $d$ is known as
'the Kondo in a box' effect.
Refs.\cite{Thimm,Simon, Baranger}
 evidentiate oscillations of $DoS_d$ due to finite size of the big dot,
and also the presence of  Kondo-type correlations which manifest itself
in an increased  spectral weight.
In our system, despite the fact that the
device is connected  to infinite leads, we find an increase of the
spectral weight above the Fermi energy (see Fig.2b). One may ask
how  the density of states looks in the case of an applied bias
larger than  the level spacing.
It is well-known that a finite bias yields the splitting of the Kondo
resonance of a single dot \cite{Meir}, however our interest concerns
the double-dot system in the strong hybridization regime. The numerical result
shows  a two-step behavior: a first increasing step occurs
 at $-V/2$ followed  by a  second one at $+V/2$ as in Fig.3.
 In the next section we shall analyze the equilibrium conductance as
 function of $V_g$.

\section{ The Fano profile of the Coulomb oscillations of the source-drain
conductance}
Our aim is to model the experimental device of Sasaki at al \cite{Sasaki}
and to analyze the output of the interplay between the Fano and Kondo
effects in the double-dot transistor-like system.
The source-drain current is the result of combined action of two channels:
i) the resonant channel passing through the big dot (controlled by
$V_g$) and ii) the Kondo channel through the small dot (controlled by
$\Gamma$ and temperature). The charge carrier may enter the big dot any
time the gate potential compensates the charging energy $\Delta=U_c +\delta E$,
where $U_c$ is due to the Coulomb repulsion  and $\delta E$ is the
distance between the levels of the big dot. In other words, the resonant
channel is activated/deactivated periodically with $V_g$ due to the
periodical blockade of the big dot.

\begin{figure}[htb]
\includegraphics[angle=-90,width=0.50\textwidth]{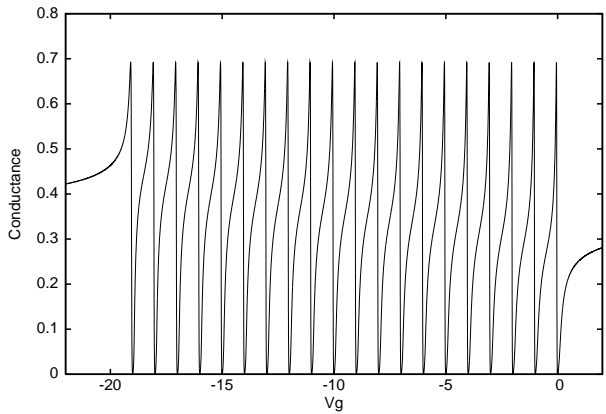}
\caption{The oscillations of the source-drain linear conductance when
the gate on the big dot $V_g$ is varied continuously. $V_g$ is measured in
units $\Delta$ and the number of levels in the big dot is $N=20$.
The small dot is in the Kondo regime: $E_d=-0.2,\Gamma=0.04$ (measured in
the leads band width), $T/T_K=1/10.$}
\end{figure}
\begin{figure}[htb]
\hskip-12cm
\includegraphics[angle=-00,width=0.80\textwidth]{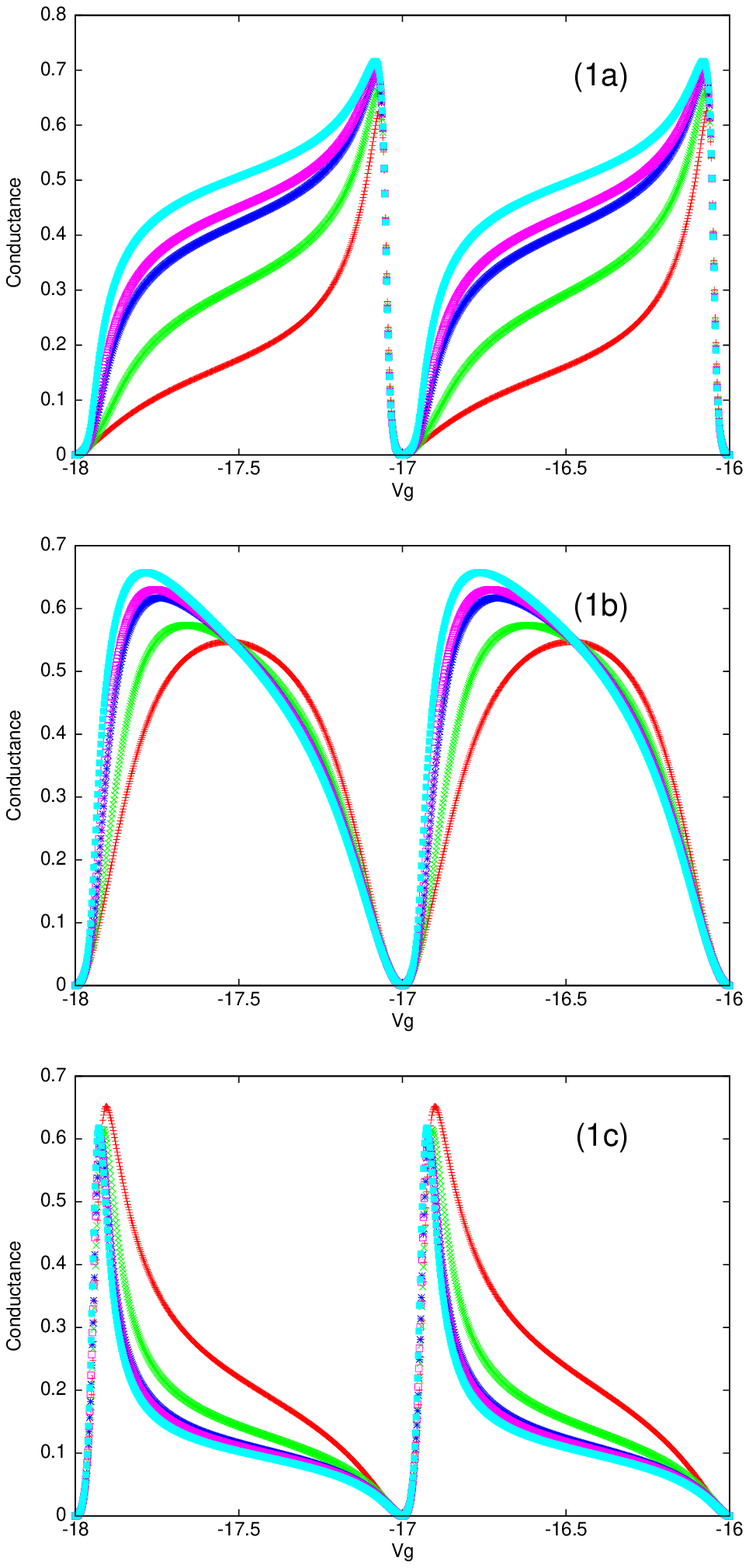}
\hskip20.0cm
\vskip-18.6cm
\includegraphics[angle=-00,width=0.80\textwidth]{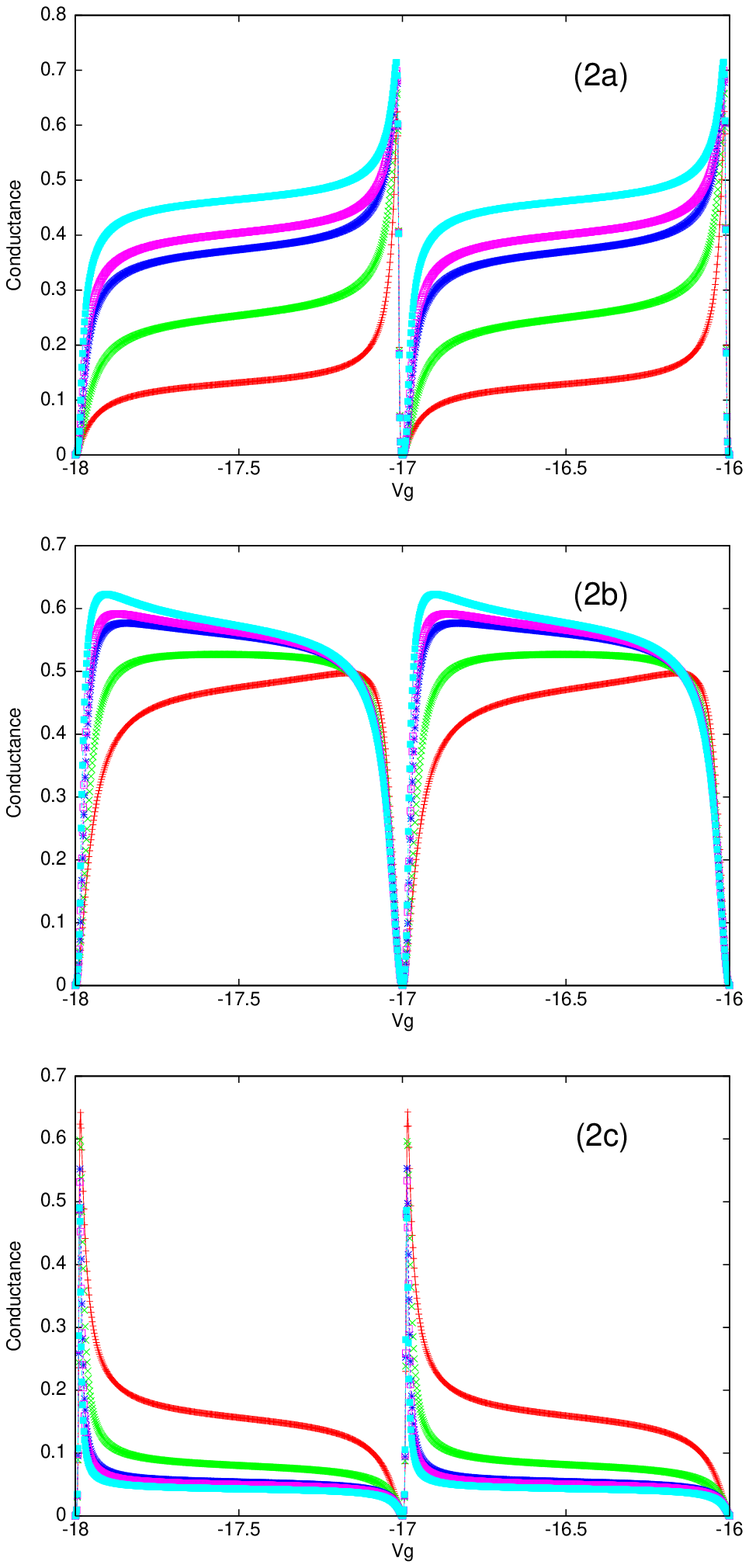}
\vskip-4.5cm
\caption{(color online) Two periods of the oscillations shown
in Fig.4 for two values
of the inter-dot hybridization $\tau=0.01$ (left) and $\tau=0.005$ (right).
At smaller $\tau$ the Fano resonance is  narrower and the background more
evident. One notices the differences in the temperature dependence of the
oscillation amplitude :  in the  Kondo regime (panels 'a') the conductance
decreases with increasing temperature, the opposite occurs in the empty
level (panels 'c'), but a visible crossing of the isotherms occurs in the
mixed valence regime (panels 'b'). (The red curves  denote the
highest temperature).}
\end{figure}
On the other hand, the device is typical for the occurrence of the Fano effect.
When the gate on the big dot is changed continuously, the Fano interference
gives rise to Fano lines which repeat in a sequence with the Coulomb
blockade periodicity $\Delta$. Denoting by $\gamma$ the width of the
Fano resonance, the ratio $\gamma/\Delta$ is responsible
for the general aspect of the sequence, as it controls the degree
of overlapping of neighboring resonances.
When this ratio is small enough the resonances are independent
and the usual Fano expression holds around each resonance:
\begin{equation}
    G(\epsilon)=const \frac{(\epsilon+q)^2}{\epsilon^2+1},
~ ~ \epsilon=\frac{E-\epsilon_0}{\gamma/2},
\end{equation}
(where $\epsilon_0$ is the resonance energy , $q$ is the Fano
asymmetry factor and the constant gives the background value).

The equilibrium source-drain conductance, for a large
range of $V_g$, is shown in Fig.4, while Fig.5 shows two periods of the
Fano oscillations for the three specific regimes of the small dot:
Kondo, mixed valence and empty level.
The two columns correspond to different values of the coupling
parameter $\tau$ which controls the resonance width. It may be observed
that small $\tau$ (the case in the right column) means a narrow
resonance, in which case the background plateau becomes more evident.

The calculation is based on eq.(3) in the limit $V=0$
and is performed at different temperatures. The upper panel evidentiates
the increase of the conductance with decreasing temperature which is the
typical effect of the Kondo correlations on the Fano lines, known as the
Fano-Kondo effect. The lowest panel describes the empty level regime of the
small dot where the temperature dependence is reversed compared to the Kondo
regime, and the shape of the curves indicates also the change of the sign
of the Fano asymmetry factor $q$. The asymmetry factor can be easily
calculated in the non-interacting case ($U_H=0$) and equals $q=E_d/\Gamma$,
indicating that the sign of $q$ is negative when the d-level lies in the
well and positive when $E_d$ is above the well.
For the interacting case an analytical expression of $q$ is not available,
however the numerical results in Fig.5 show that
the same change of sign  holds when moving from the Kondo to empty level
regime.

The crossover between the two extreme situations is shown in the middle panel
which describes the mixed-valence regime. The significant issue is the presence
of crossing points of the isotherms where $dG/dT=0$. The most visible
crossing occurs in the panel (1b) : one notices that to the right of the
crossing point
the conductance increases with the temperature (as in the empty
level case), while to the left the dependence is opposite
(similar to the Kondo regime). Such crossing points are present also in
the cases (a) and (c), but they are very close to the Fano zero,
and are not visible in Fig.5.

The analytical expressions of the conductance (extended formulae for
complex meso-systems in the Lacroix's approximation \cite{Lacroix} are given in
\cite{Meir05}) are too cumbersome to evidentiate the crossing points, so
we have to confine ourselves to  numerical studies.
However, because of the competition between the interference and Kondo effect,
the change of the temperature behavior is understandable
 for the double dot device.
Indeed, although the Kondo channel is reinforced with
decreasing temperature, its superposition with the resonant channel may
give rise to a destructive interference which results in a reduction of
the conductance when the temperature is decreased.
A similar property has been found for the triangle interferometer with
magnetic impurity \cite{DTA}.

Finally, we note that if $V_g$ applied on the big dot is an a.c. voltage of
frequency $\Omega_G$,
the source-drain current oscillates with a higher frequency $\Omega_{sd}=
N_g \Omega_G$, where $N_g$ is the number of levels in the big dot which are scanned during one period. In technical terms, one may say that the lateral
double-dot device works as a voltage-controlled transistor oscillator, with the
property of frequency multiplication.

\section{ Non-equilibrium conductance: zero-bias anomalies and
scaling properties in the Kondo regime}

It is known that a bias $V$ applied  on a single quantum dot in the Kondo regime
suppresses the correlations \cite{WM,Rosch} so that the largest value of
the differential
conductance occurs at $V=0$, or in other words, the differential conductance
$G(V)=dI/dV$ looks peak-like as function of the bias.
In what concerns the side-coupled double-dot system,
one expects a peak-like behavior for those values of the  gate potential $V_g$ and temperature $T$, where the
system behaves Kondo-like. However we have already mentioned
that in some range of $\{V_g,T\}$ the interference
strongly compete the correlations. In such a situation, it is reasonable
to expect
a modified
behavior of $dI/dV$, and a dip-like aspect (i.e.,the enhancement with
the increasing bias) cannot be {\it a priori} excluded. In what follows we shall
give  plausibility arguments which, together with the numerical results,
indicate the presence of a peak-dip crossover of the differential conductance.

Let $t_r$ and $t_c$ be the tunneling amplitude along the two channels and
$t$ the amplitude of the total transmittance through the device:
\begin{equation}
    |t|^2=|t_c+t_r|^2=|t_c|^2+|t_r|^2+2|t_c| |t_r|cos(\phi_c-\phi_r)
\end{equation}
The transmittance $|t_c|$ along the continuous (Kondo) path is controlled by
$V$ and $T$, both parameters killing the correlations at large values, so that
at $T<T_K$, one has $d|t_c|/dV <0$ and also $d|t_c|/dT <0$.
The peak or dip aspect effect  is given by the sign of the derivative:
\begin{equation}
\frac{d|t|^2}{dV}=2 \frac{d|t_c|}{dV}~[|t_c|+|t_r|cos(\phi_c-\phi_r)]
\end{equation}
Since the sign of $\frac{d|t_c|}{dV}$ is fixed, the enhancement or depletion
of the total transmittance with increasing bias depends on the sign of the
quantity $A=|t_c|+|t_r|cos(\phi_c-\phi_r)$.

\begin{figure}[htb]
\includegraphics[angle=-90,width=0.5\textwidth]{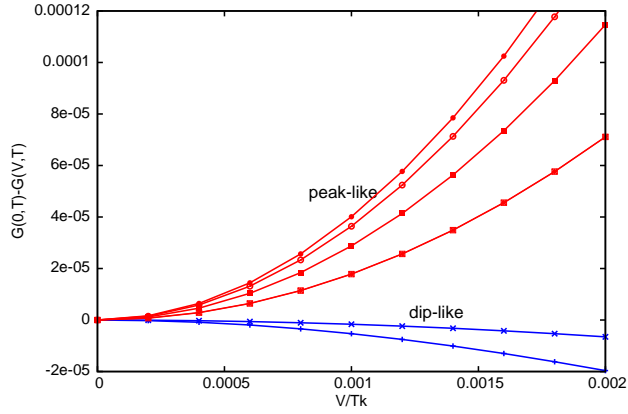}
\caption{(color online)
$\Delta G(V,T)$ as function of the source-drain bias (scaled with $T_K$)
showing the dip-peak crossover when the
gate potential is changed from $V_g/\Delta=-17.00$ (lowest curve) to
$V_g/\Delta=-17.70$
(highest curve) at fixed temperature $T=T_K/100$. The small dot is in
the Kondo regime $\Gamma/|E_d|=0.2$ .}
\end{figure}
\begin{figure}[htb]
\includegraphics[angle=-90,width=0.5\textwidth]{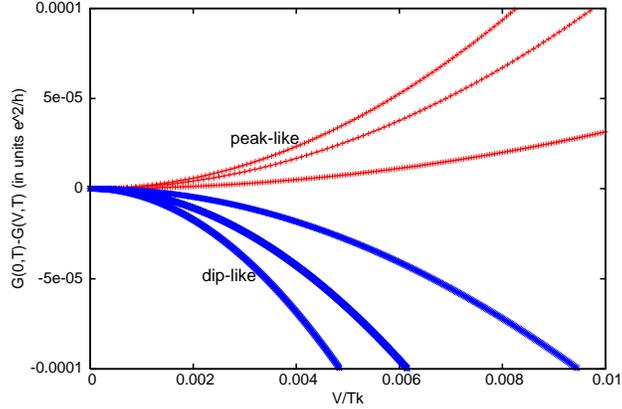}
\caption{(color online)
$\Delta G(V,T)$ as function of the source-drain bias
(scaled with $T_K$)
showing the dip-peak crossover with increasing temperature at fixed
gate potential $V_g/\Delta=-17.10$. The lowest curve correspond to $T/T_K=1/80$
the highest curve to $T/T_K=1/5$. $\Gamma/|E_d|=0.4$ .}
\end{figure}

The above  expression points out that the dip-like dependence of the
transmittance as function of the bias is allowed for values of
$\{V_g,T\}$ such that $A<0$, fact that expresses the implication of
phases in this effect.
Indeed, the numerical calculations show that the zero bias anomaly may
exhibit
not only the enhancement but also a depletion of differential conductance
depending on the interference conditions. Fig.6 depicts the quantity
$\Delta G(V,T)=G(0,T)-G(V,T)$ as function of $V$
for different values
of the gate potential $V_g$ at a given temperature. Depending on $V_g$
both the decrease and increase of the source-drain bias can be noticed.

Let us now keep a fixed gate and use the temperature as the
control parameter. This case is exemplified in  Fig.7 which shows that
also in this situation the conductance  undergoes a crossover from dip
to peak behavior: the dips occur at low $T$, and with increasing
temperature the
conductance dependence on $V$ becomes peak-like.

With the ansatz that $\Delta G (V,T,T_k)$ is a homogeneous function,
scaling properties of this function have been studied. For the fully
screened Kondo problem of the single dot,
Ralph {\it et al.} \cite{Ralph}  suggest that  $\Delta G$ scales like:
\begin{equation}
    \Delta G(V,T) \sim T^{s_1} F(VT^{s_2})
\end{equation}
with the exponents $s_1=2 ,s_2=-1 $. A deviation from this law has been
observed already in the limit of low-V and low-T by Schiller and
Herschfield \cite{Schiller}.
Grobis et al \cite{Grobis} describe their experimental data
in terms of scaled variables $V/T_K$ and $T/T_K$ by the
following expression (eq.(4) in \cite{Grobis}):
\begin{equation}
\frac{  \Delta G(V,T)}{c_T G_0} \approx \alpha \big(\frac{V}{T_K}\big)^2
\big[1- \frac{c_T}{\alpha\gamma} \big(\frac{T}{T_K}\big)^2\big] ,
\end{equation}
with universal coefficients along the Kondo plateau, slight variations
being observed however in the mixed valence regime.
The coefficient $\alpha$ has been  calculated also theoretically in
Refs.\cite{Sela,Mora}.

\begin{figure}[htb]
\includegraphics[angle=-00,width=0.9\textwidth]{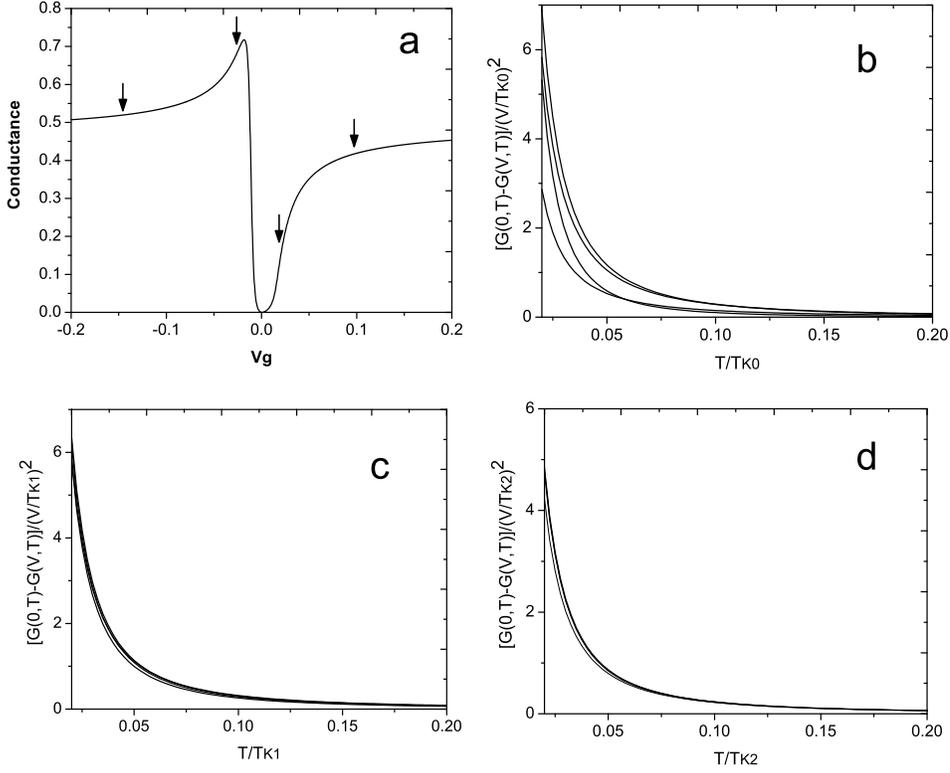}
\vskip-0.5cm
\caption{
 a) A Fano-Kondo line. The arrows indicate
four values of $Vg$ (form left to right: -0.15, -0.03, 0.03 and 0.1)
for which $\Delta G$ is plotted. Three definitions are
considered for the Kondo temperature:
 b) The SIAM Kondo temperature $T_{K0}$, which does not depend on $Vg$,
c) The half-width of the Kondo peak at $T=0$, as estimated in
\cite{Meir05} - $T_{K1}$.
 d) The half-width of the Kondo
peak at a finite temperature, calculated as described in text -
$T_{K2}$. In both cases c) and d) the Kondo temperature depends on
$Vg$, and the scaling is satisfactory, resulting in the data
collapse. For all curves, $V=T_K/100.$}
\end{figure}

We look for similar properties in the Fano-Kondo regime of the
double dot system. The problem becomes more difficult since,
besides the variables V, T and the parameters describing the Kondo
dot (packed  into the Kondo temperature $T_{K0}$ of the SIAM),
the function $\Delta G$ depends also on the parameters that describe
the state of the side-dot. It is obvious that the SIAM Kondo
temperature $T_{K0}$ is no longer a suitable scaling parameter. We need a
scaling parameter -an effective Kondo temperature $T_K$- that accounts also
for the spectral influence of the side-coupled dot. This Kondo temperature will
depend on  $V_g$. In addition, since the system may shift from peak to dip
behavior (as proved in Figs.6-7), it seems impossible to find a unique scaling
 law describing both the peak and dip regions.
It remains however a legitimate task to seek scaling properties in the
regions with Kondo behavior along the Fano line. In Fig.8a the arrows indicate
four values of the gate $V_g$ for which we test the scaling law (in all
these points the non-equilibrium conductance behaves peak-like; the dip-type
dependence occurs only in a narrow vicinity of the Fano zero). In other words,
we check whether the function $\Delta G (V,T)$ may be scaled in terms of the
variables $V/T_K$ and $ T/T_K$. A quadratic dependence on the bias can
be assumed from the start, as we work in the limit of low bias and
therefore it is  legitimate to keep only the quadratic term.
The $V^2$ dependence of $\Delta G$ is visible also in Figs.6-7.

Fig.8 shows the function $\Delta G/(V/T_K)^2$ plotted versus
$T/T_K$ for different definitions of $T_K$ in order to see if the
data collapse occurs. In Fig.8b we use $T_{K0}$ (the SIAM Kondo
temperature) and, as expected, the collapse of the curves
(corresponding to the four values of $V_g$) fails.

Then we check two
other possible definitions for the effective $T_K$. The first
(referred as $T_{K1}$) is that one proposed in \cite{Meir05} where
$T_{K1}$ was calculated as the half-width of the Kondo resonance at
$T=0$ and deep level regime. Using simplifying assumption, the
authors obtained an analytical expression (eq.60). By using $T_{K1}$,
our curves fall on top of each other with a precision of a few
percents. The situation is plotted in Fig.8c.

Keeping in mind the difficulties to reach numerically the $T=0$ limit, we
calculate also the half-width of the Kondo resonance at
a low but finite temperature T=Tk/100 (which we call $T_{K2}$ \cite{tk2}).
 This proves also a suitable scaling parameter, leading to the data collapse in
Fig.8d.

Two comments are necessary:

1. We have looked for the scaling
parameter in two different ways corresponding in fact to two different
estimates of the Kondo temperature defined as the  half-width of the
Kondo resonance in DoS.
The important thing is that  $T_{K1}$ and $T_{K2}$ vary {\it in the same
way} at the change of the system parameters, and the ratio of the two
remains roughly constant $T_{K2}/T_{K1}\in [0.11,0.14]$.

2. The EOM is an approximate method and
this may be responsible for the slight imprecision in the
fit of the scaled curves,
and also for shape of the scaling function which has the correct
decreasing behavior, but is not quadratic as predicted by
\cite{Schiller} (however, that prediction regards the SIAM, and not the
Fano-Kondo set-up).

\section{Conclusions}

Recently, Sasaki {\it et.al.}\cite{Sasaki} proposed a new
set-up for the study of the Fano-Kondo effect: a Kondo dot is
directly coupled to conducting leads and a second dot (assumed
non-interacting) is side-coupled to the Kondo dot.
The set-up mimics a transistor, where the
side-dot plays the role of the basis, and whose functionality
is controlled by the level of the Kondo correlations, but also by the
interference processes.

Due to the many-level aspect of the big dot, the
source-drain current presents oscillations with Fano profile when the gate
is varied continuously.
In the Kondo regime, the Fano lines become more asymmetric as the temperature
decreases (in concordance with \cite{Sasaki}) corresponding to the increase
of the Kondo direct transmission. The Fano dip always reaches {\it zero}
conductance, as resulting also with the slave boson method \cite{Tamura,Liu}.

The new physics is interesting since the strong hybridization of the dots
has a significant influence on the Kondo effect.
In the vicinity of the Fano zero, the Kondo effect is strongly
suppressed and the conductance varies negligibly with the
temperature. Moreover, in this regime, the destructive interference leads to
a  {\it zero} bias dip-type  anomaly  in the differential conductance.
Outside this region, the differential conductance shows the expected peak
behavior.

 Interestingly, the non-equilibrium  conductance
exhibits scaling properties similar to the  SIAM, even in the presence
of the Fano interference. Nevertheless, the SIAM Kondo temperature is no
more the suitable scaling parameter.
Provided  we are in the low bias and low temperature domain,
the  function $\Delta G = G(0,T)-G(V,T)$
calculated for different values of $V_g$,
depends  {\it only} on $V/T_K$ and $T/T_K$, where the scaling invariant $T_K$
is extracted from the half-width of the Kondo peak in $DoS_d$ of the double dot.

We have used the Keldysh transport formalism and the
equation of motion method which captures not only the Kondo spin
fluctuations but also the charge fluctuations, an important ingredient for
the aspect of the Fano-Kondo resonances.

In a very recent paper, Balseiro {\it et. al.} \cite{Balseiro} also show
that the EOM method is
suitable for checking scaling properties in the Kondo regime.
However they address the SIAM, and not the Fano-Kondo set-up.

\section{Acknowledgements}
We acknowledge  support from PNCDI2-Research Programme
(grant no 515/2009), Core Programme (contract no.45N/2009) and
Sonderforschungsbereich 608 at the  Institute of Theoretical Physics,
University of Cologne.
One of the authors (AA) is very much indebted to R.Bulla and E.Sela
for illuminating discussions.


\begin{references}
\bibitem{tk2}$T_{K2}$ is calculated by solving  numerically the
 equation $ImG(Ef+T_{K2}/2)=1/2 ImG(Ef)$ , at
 T=TK2/100, after the background is removed (the background is the value of
the density of states at a high temperature, when the Kondo peak is absent).
\end{references}
\end{document}